\documentclass[twocolumn,tighten,times,numberedappendix,twocolappendix]{aastex631}
\usepackage{CJK}
\usepackage{natbib}

\usepackage{xcolor}
\usepackage{multirow}
\usepackage{booktabs}

\received{\today}

\submitjournal{ApJ}

\begin{document}

\title{\large An XMM-Newton view of the Symbiotic Stars HM\,Sge, NQ\,Gem, and PU\,Vul}

\correspondingauthor{Jes\'{u}s A. Toal\'{a}}
\email{j.toala@irya.unam.mx}

\begin{CJK*}{UTF8}{gbsn}

\author[0000-0002-5406-0813]{Jes\'{u}s~A.~Toal\'{a}\,(杜宇君)}
\affil{Instituto de Radioastronom\'{i}a y Astrof\'{i}sica, UNAM, Ant. carretera a P\'{a}tzcuaro 8701, Ex-Hda. San Jos\'{e} de la Huerta, 58089 Morelia, Mich., Mexico}

\author[0000-0002-7173-5077]{Marissa~K.~Botello}
\affil{Instituto de Astronom\'{i}a, Universidad Nacional Aut\'{o}noma de M\'{e}xico, Apdo. Postal 877, 22860 Ensenada, B.C., Mexico}

\author[0000-0003-0242-0044]{Laurence~Sabin}
\affil{Instituto de Astronom\'{i}a, Universidad Nacional Aut\'{o}noma de M\'{e}xico, Apdo. Postal 877, 22860 Ensenada, B.C., Mexico}



\begin{abstract}
We present the analysis of archival XMM-Newton observations of the symbiotic stars HM~Sge, NQ~Gem, and PU\,Vul. The EPIC-pn spectra hint at the presence of emission lines, which are further confirmed in the 1st order RGS spectra of the three sources. Spectral modeling of the EPIC-pn data disclose unprecedented characteristics, for instance, the best fit to the EPIC-pn spectrum of the $\beta$-type symbiotic star PU~Vul reveals the presence of two plasma components. We report the discovery of an extremely soft spectral component in the EPIC-pn spectrum of the $\beta$-type symbiotic star HM~Sge which we suggest is produced by periodic mass ejections such as jets. Consequently, we suggest that a simple $\beta$-type classification no longer applies to HM~Sge. Finally, the spectrum of the $\beta/\delta$-type symbiotic star NQ~Gem can not be fitted by a two-temperature plasma model as performed by previous authors. The model requires extra components to fit the 1.0--4.0~keV energy range. More sophisticated models to $\beta/\delta$-type symbiotic stars are needed in order to peer into the accretion process from such systems.  
\end{abstract}

\keywords{\href{http://astrothesaurus.org/uat/1674}{Symbiotic binary stars (1674)};
\href{http://astrothesaurus.org/uat/2050}{Stellar accretion(1578)};
\href{https://astrothesaurus.org/uat/1799}{White dwarf stars(1799)};
\href{http://astrothesaurus.org/uat/2050}{X-ray stars};
\href{http://astrothesaurus.org/uat/2050}{Low mass stars(2050)}
\vspace{4pt}
\newline
}


\section{Introduction} \label{sec:intro}
\end{CJK*}
\label{sec:intro}

Symbiotic stars are binary systems composed of a compact object accreting enough material from a red giant to produce observable emission at any wavelength. This definition, proposed by \citet[][]{Luna2013}, is intended to be as free as possible from observational selection biases.
The definition is somewhat loose if one recognizes that the compact object might be a white dwarf (WD), a neutron star, or even a black hole \citep[see][]{Luna2013}. For this, symbiotic systems have been divided into WD symbiotics and symbiotic X-ray binaries \citep[see for example][]{Chakrabarty1997,Enoto2014,Hinkle2019}. In this paper, we will refer to WD symbiotics simply as symbiotic stars.

In symbiotic stars the WD accretes material from the red giant companion, but it is currently not clear whether the WD accretes material through a Bondi-Hoyle process \citep{BondiHoyle1944}, by Roche-lobe overflow, or a hybrid wind Roche-lobe overflow channel \citep[see][]{Podsiadlowski2007}. Nevertheless, an accretion disk forms surrounding the compact object.

\begin{table*}
\caption{XMM-Newton observations of the three symbiotic stars analyzed here. All observations were obtained with the full-frame mode and the medium optical blocking filter.}
\setlength{\tabcolsep}{0.95\tabcolsep}  
\label{tab:obs}
\begin{center}
\begin{tabular}{cccccccccccc}
\hline
Revolution & Obs.~ID. & Object & Observation start & Observation end     & \multicolumn{3}{c}{Total exposure time} & \multicolumn{3}{c}{Useful exposure time}\\
\cmidrule(lr){6-8}\cmidrule(lr){9-11}
       &             &        &                   &                     & pn      & MOS1   & MOS2 & pn & MOS1 & MOS2\\
       &             &        &  (UTC)            &  (UTC)              & (ks)    & (ks)   & (ks) & (ks)    & (ks)   & (ks)\\
\hline
2725 & 0740610101 & NQ~Gem  & 2014-10-26T12:35:57 & 2014-10-27T06:17:37 & 60.75  & 62.35 & 62.32 & 44.80 & 57.72 & 56.35 \\      
3010 & 0784910101 & PU\,Vul & 2016-05-16T20:56:34 & 2016-05-17T15:14:54 & 62.21  & 60.66 & 60.64 & 24.69 & 41.68 & 36.55\\ 
3078 & 0784910201 & HM~Sge  & 2016-09-29T16:51:07 & 2016-09-29T20:27:47 & 10.04  & 11.65 & 11.63 & 7.76 & 11.20 & 10.58 \\
\hline
\end{tabular}
\end{center}
\end{table*}

X-ray observations have been classically used to assess the accretion process in WD symbiotic stars, but not many  have been detected in X-rays. There are less than 300 confirmed galactic WD symbiotic stars reported in the New Online Database of Symbiotic Variables by 2023 January 30\footnote{\url{http://astronomy.science.upjs.sk/symbiotics/}} \citep[see also][]{Akras2019,Merc2019a}, but \citet{Merc2019} reported that only about 60 have been detected with X-ray instruments. \citet{Merc2019} suggested that the production of X-ray emission from symbiotic stars should be a common feature, but thus far, the X-ray detections mostly correlate with the brightest and closer systems.

The production of X-rays from symbiotic stars can be attributed to different origins \citep[see][and references therein]{Mukai2017}. Extremely soft X-rays ($E<$0.5~keV) can be produced by nuclear burning at the surface of the WD. X-rays are also produced when the accreted material hits the surface of the WD \citep[see the early work of][]{Aizu1973}, but details depend whether the WD is magnetically active or not. In addition, outflows originated by the central engine can also produce shocks that heat up the material up to X-ray-emitting temperatures. Strong shocks can be produced either by the fast wind from the WD component or jet-like ejections, both interacting with the slower wind from the cooler companion. High-quality images and spectra are most needed in order to put into context this phenomenology in symbiotic stars \citep[see, e.g.,][]{Karovska2010}.

The first X-ray classification of symbiotic stars was presented by \citet{Murset1997} dividing these objects into three different groups depending on their spectra properties. $\alpha$-type symbiotic stars corresponds to those sources with super soft X-ray spectra peaking at energies bellow 0.4~keV and attributed to the quasi-steady thermonuclear burning on the surface of the WD \citep[see][]{Orio2007}. $\beta$-type objects with spectral peaks close to 0.8~keV are associated with optically-thin plasma with temperatures $\sim10^6$~K produced by colliding winds, accretion shocks and/or accretion disk. Finally, $\gamma$-type corresponds to harder X-ray sources with emission associated with shocks between the accretion material of the compact object that reaches up to 2.4 keV energy. We note that this classification scheme was defined by using ROSAT data which was sensitive to the soft X-ray emission in the 0.1--2.5 keV energy range.

However, given that some symbiotic stars were discovered to emit X-rays with energies above 20~keV \citep[e.g.,][]{Chernyakova2005,Kennea2009}, the original classification scheme proposed by \citet{Murset1997} had to be revised. \citet{Luna2013} expanded the $\alpha/\beta/\gamma$ classification scheme by adding a fourth category, $\delta$-type for highly absorbed, hard X-ray emission sources. These authors used Swift XRT observations from symbiotic stars covering the 0.4--10~keV energy range. In particular, the X-ray emission from these sources is typically highly extinguished and very likely corresponds to emission from the innermost accretion region surrounding the WD component. \citet{Luna2013} demonstrated that symbiotic stars not only emit soft X-rays, but there was a significant population of objects with hard X-ray emission.

In this work we analyze publicly available, unpublished XMM-Newton observations of the symbiotic stars HM\,Sge, NQ\,Gem, and PU~Vul which have been classified as $\beta$-, $\beta/\delta$-, and $\beta$-type, respectively \citep[see][and references therein]{Merc2019}. We note that HM~Sge and PU\,Vul have not been studied since the work of \citet{Murset1997} but NQ~Gem is one of the symbiotic stars demonstrated to be part of the $\beta/\delta$-type using Swift data by \citet{Luna2013}.
The analysis of the XMM-Newton data presented here improves previous spectral characterization of these three symbiotic stars. This paper is organized as follows. In Section~\ref{sec:obs} we present the observations and their preparation. Section~\ref{sec:results} presents our results and their discussion. Finally, in Sections~\ref{sec:summary} we list our conclusions.

\begin{figure*}
\begin{center}
\includegraphics[angle=0,width=\linewidth]{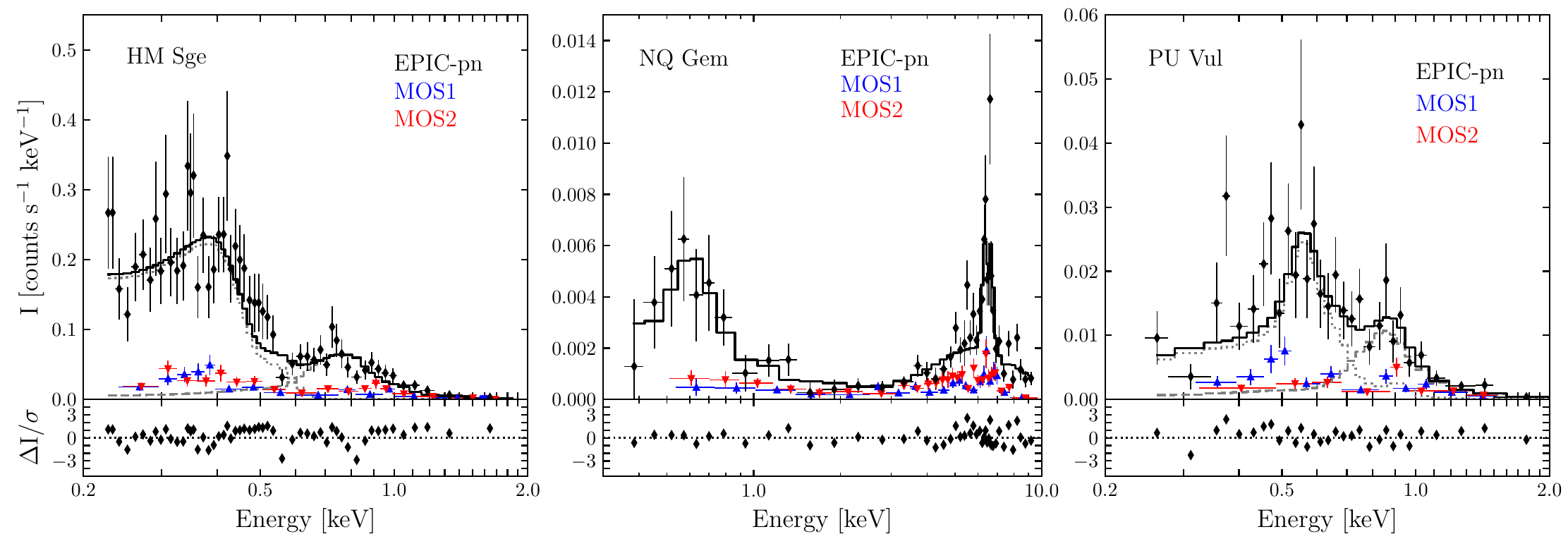}
\caption{Background-subtracted XMM-Newton EPIC spectra of HM\,Sge (left), NQ~Gem (middle), and PU~Vul (right). Colored symbols correspond to the observed spectra while the black histograms represent the best-fit models to the EPIC-pn data. The (gray) dotted and (dashed) lines in the HM~Sge and PU~~Vul panels represent the two different plasma components of their best fit models described in Table~\ref{tab:model}. In all cases the bottom panels show the residuals.}
\label{fig:spec1}
\end{center}
\end{figure*}

\section{Observations and data preparation}
\label{sec:obs}

HM\,Sge, NQ\,Gem, and PU\,Vul were observed by XMM-Newton with the European Photon Imaging Cameras (EPIC) on different seasons. In Table~\ref{tab:obs} we list the details of each observation set. The observations data files were obtained from the XMM-Newton Science Archive\footnote{\url{http://nxsa.esac.esa.int/nxsa-web/\#search}}. In all cases the three EPIC cameras (pn, MOS1, and MOS2) were used in the full-frame mode with the medium optical blocking filter. The total observing times for HM\,Sge, NQ\,Gem, and PU\,Vul are 13.0, 63.7, and 65.0, respectively. The data were processed with the Science Analysis Software \citep[SAS, version 20.0;][]{Gabriel_2004} with the calibration files obtained on 2022 August 19.

Bad periods characterized by high-background levels were evaluated by analyzing EPIC light curves extracted in the 10.0--12.0~keV energy range. Typical values of 0.5 and 0.2~counts~s$^{-1}$ were set as maxima for the pn and MOS cameras, respectively. After cleaning the data, we extracted pn, MOS1, and MOS2 spectra for the three symbiotic stars using the SAS task {\it evselect} adopting circular apertures centered on each source with radii of 20$''$. For the MOS observations events with CCD patterns 0--12 were selected, but for the EPIC-pn data, only events with CCD pattern 0 (single pixel events) were selected. The background spectra were extracted using adjacent regions with no contribution from the symbiotic star or any other source. The calibration matrices were produced with the {\it rmfgen} and {\it arfgen} SAS tasks. The resultant background-subtracted spectra of the three symbiotic stars are presented in Fig.~\ref{fig:spec1}.

In addition, we also processed the Reflection Grating Spectrometers (RGS) data also on board XMM-Newton. Although the quality of the RGS spectra hampers a more comprehensive analysis, these were inspected to search for the presence of emission lines which helped us achieve a better interpretation of the EPIC spectra. Some details of the analysis of RGS data as well as the resultant spectra are presented in Appendix~\ref{sec:appendix}.

The EPIC spectra were modeled with the X-Ray Spectral Fitting Package \citep[XSPEC; version 12.12.1;][]{Arnaud1996}. Extinction of X-rays caused by the interstellar medium (ISM) was included by adopting the Tuebingen-Boulder absorption model {\it tbabs} \citep{Wilms2000} which takes into account the absorption produced by the gas-phase ISM, the grain-phase ISM, and the molecules in the ISM\footnote{See further details in \url{https://heasarc.gsfc.nasa.gov/xanadu/xspec/manual/node268.html}}. The hydrogen absorption column densities ($N_\mathrm{H}$) for the three symbiotic stars analyzed here were fixed to the values reporte by the NASA's HEASARC $N_\mathrm{H}$ column density tool \citep{HI4PI2016,Kalberla2005,Dickey1990}\footnote{\url{https://heasarc.gsfc.nasa.gov/cgi-bin/Tools/w3nh/w3nh.pl}}. We adopted the distances estimated by \citet{BailerJones2021} using Gaia data (see Table~\ref{tab:model}).

In order to fit the EPIC spectra of the three symbiotic stars, we used different model components or a combination of them. Given that symbiotic stars emit X-rays through thermal processes \citep[see][]{Mukai2017} most of our spectral fits were performed adopting optically-thin collisionally-ionized emission plasma models. In this paper we use the {\it apec} emission spectrum\footnote{\url{https://heasarc.gsfc.nasa.gov/xanadu/xspec/manual/XSmodelApec.html}} included in XSPEC. In all cases we adopted the solar abundances from \citet{Lodders2009}. We were able to extract spectra from the three EPIC cameras (pn, MOS1, and MOS2) as illustrated in Fig.~\ref{fig:spec1}. All EPIC spectra were inspected, but the EPIC-pn spectra have superior count rates than those extracted from the MOS instruments. Thus, the spectral modeling will be only performed for those spectra. Nevertheless, we note that the analysis of the MOS spectra result in similar models as those obtained from the EPIC-pn spectra.

Two-temperature plasma components were sufficient to produce good fits to the X-ray emission from HM~Sge and PU~Vul, but a more complex model was needed for NQ~Gem \citep[see][]{Luna2013}. The best models are described in detail in the following section. The goodness of the model fits was assessed by the reduced chi squared statistics ($\chi_\mathrm{DoF}^2$), defined as the $\chi^2$ statistics per degree of freedom (DoF). These values are provided in XSPEC and are also listed in Table~\ref{tab:model} for each source.

\begin{table*}
\caption{Model parameters of the best fits to the EPIC-pn spectra. The net time ($t_\mathrm{net}$), count rate, and total number of counts correspond to the EPIC-pn data of each source. $f_\mathrm{X}$ and $F_\mathrm{X}$ are the observed and intrinsic fluxes, respectively, whilst $L_\mathrm{X}$ is the luminosity.} 
\setlength{\tabcolsep}{\tabcolsep}  
\label{tab:model}
\begin{center}
\begin{tabular}{lccc}
\hline
                            & HM\,Sge                    & NQ\,Gem                        & PU Vul                  \\
                            &                            &                                &                         \\
\hline
$d$ [kpc]                   & 1.0                        &  1.04                          & 4.1                     \\
$t_\mathrm{net}$ [ks]       & 7.8                        & 44.8                           & 24.7                    \\
Count rate [cnts~s$^{-1}$]  & 8.97$\times10^{-2}$        & 1.65$\times10^{-2}$            & 1.22$\times10^{-2}$     \\
Total counts [cnts]         & 700                        & 740                            & 300                     \\
Energy range [keV]    & 0.2--2.0                   & 0.3--9.0                       &0.2--2.0                 \\
\hline
$\chi_\mathrm{DoF}^{2}$     & 62.31/51=1.21              & 36.20/32=1.13                  & 32.99/26=1.27           \\
$N_\mathrm{H,1}$ [$10^{21}$~cm$^{-2}]$    & 3.0          & 0.49                           & 1.7                     \\
$kT_{1}$ [keV]     & (2.9$^{+0.2}_{-0.2}$)$\times$10$^{-2}$ & 0.22$^{+0.03}_{-0.03}$      & 0.15$^{+0.04}_{-0.03}$  \\
$A_{1}$  [cm$^{-5}$]        & 122                        & (2.6$\pm$0.8)$\times10^{-6}$             & (3.6$\pm$0.9)$\times10^{-5}$ \\
$f_1$ [erg~cm$^{-2}$~s$^{-1}$] & (9.8$\pm$3.4)$\times$10$^{-14}$ & (2.9$\pm$1.1)$\times$10$^{-15}$  & (9.3$\pm$2.1)$\times$10$^{-15}$\\
$F_1$ [erg~cm$^{-2}$~s$^{-1}$] & (5.3$\pm$1.9)$\times$10$^{-10}$ & (4.0$\pm$1.9)$\times$10$^{-15}$  & (6.8$\pm$1.6)$\times$10$^{-14}$\\
$L_1$ [erg~s$^{-1}$]           & (6.3$\pm$2.2)$\times$10$^{34}$  & (5.2$\pm$2.6)$\times$10$^{29}$   & (1.4$\pm$0.3)$\times$10$^{32}$\\
$kT_{2}$ [keV]              & 0.31$^{+0.07}_{-0.04}$     & 2.5$^{+3.0}_{-2.0}$            & 0.75$^{+0.29}_{-0.16}$  \\
$A_2$    [cm$^{-5}$]        & 8.2$\times10^{-5}$         & (3.0$\pm$0.3)$\times10^{-6}$             & (5.1$\pm$0.9)$\times10^{-6}$ \\
$f_2$ [erg~cm$^{-2}$~s$^{-1}$] & (4.4$\pm$0.8)$\times$10$^{-14}$ & (3.1$\pm$1.3)$\times$10$^{-15}$  & (7.6$\pm$1.3)$\times$10$^{-15}$\\
$F_2$ [erg~cm$^{-2}$~s$^{-1}$] & (1.7$\pm$0.3)$\times$10$^{-13}$ & (3.5$\pm$1.4)$\times$10$^{-15}$  & (1.3$\pm$0.2)$\times$10$^{-14}$\\
$L_2$ [erg~s$^{-1}$]           & (2.0$\pm$0.4)$\times$10$^{31}$ & (4.5$\pm$1.8)$\times$10$^{29}$    & (2.6$\pm$0.5)$\times$10$^{31}$\\
$\Gamma$                    & \dots       & $-$1.5$\pm$0.5               & \dots                   \\
$A_\mathrm{pow}$ [cm$^{-5}$]& \dots       & (1.2$\pm$1.4)$\times10^{-7}$ & \dots                   \\
$f_\mathrm{pow}$ [erg~cm$^{-2}$~s$^{-1}$] & \dots & (1.2$\pm$0.1)$\times$10$^{-13}$ & \dots \\
$F_\mathrm{pow}$ [erg~cm$^{-2}$~s$^{-1}$] & \dots & (1.2$\pm$0.1)$\times$10$^{-13}$ & \dots \\
$L_\mathrm{pow}$ [erg~s$^{-1}$]           & \dots & (1.5$\pm$0.2)$\times$10$^{31}$  & \dots \\
\hline
$N_\mathrm{H,2}$ [$10^{21}$~cm$^{-2}]$   & \dots         & 390$\pm$170                    & \dots                   \\
$kT_3$  [keV]               & \dots                      & 4.5$\pm$2.0                    & \dots                   \\
$A_3$   [cm$^{-5}$]         & \dots                      & (4.8$\pm$1.1)$\times10^{-4}$   & \dots                   \\
$f_3$ [erg~cm$^{-2}$~s$^{-1}$] & \dots & (1.1$\pm$0.2)$\times$10$^{-13}$ & \dots \\
$F_3$ [erg~cm$^{-2}$~s$^{-1}$] & \dots & (1.1$\pm$0.2)$\times$10$^{-13}$ & \dots \\
$L_3$ [erg~s$^{-1}$]           & \dots & (1.4$\pm$0.3)$\times$10$^{31}$ & \dots \\
$E_\mathrm{line}$ [keV]     & \dots                      & 6.4$\pm$0.1                    & \dots                   \\
$\sigma_\mathrm{line}$[keV] & \dots                      & (1.1$\pm$0.1)$\times10^{-1}$   & \dots                   \\
$A_\mathrm{line}$ [cm$^{-5}$] & \dots                    & (5.6$\pm$1.3)$\times10^{-6}$   & \dots                   \\
$f_\mathrm{line}$ [erg~cm$^{-2}$~s$^{-1}$] & \dots & (2.8$\pm$0.5)$\times$10$^{-14}$ & \dots \\
$F_\mathrm{line}$ [erg~cm$^{-2}$~s$^{-1}$] & \dots & (2.8$\pm$0.5)$\times$10$^{-14}$ & \dots \\
$L_\mathrm{line}$ [erg~s$^{-1}$] & \dots  & (3.6$\pm$0.6)$\times$10$^{30}$ & \dots \\
\hline
$f_\mathrm{X}$ [erg~cm$^{-2}$~s$^{-1}$]  & (1.4$\pm$0.5)$\times$10$^{-13}$  & (2.7$\pm$0.5)$\times10^{-13}$  & (1.7$\pm$0.4)$\times10^{-14}$  \\
$F_\mathrm{X}$ [erg~cm$^{-2}$~s$^{-1}$]  & (5.3$\pm$2.1)$\times$10$^{-10}$  & (1.0$\pm$0.2)$\times10^{-12}$  & (8.2$\pm$1.8)$\times10^{-14}$  \\
$L_\mathrm{X}$ [erg~s$^{-1}$]            & (6.3$\pm$2.5)$\times$10$^{34}$   & (1.3$\pm$0.3)$\times$10$^{32}$ & (1.6$\pm$0.4)$\times$10$^{32}$ \\
\hline
\end{tabular}
\end{center}
\end{table*}

\section{Results and discussion}
\label{sec:results}

The analysis of the EPIC-pn spectra of HM\,Sge, NQ~Gem, and PU~Vul resulted in different models with different implications. Consequently, we will present their results and discussion in separate subsections. Details of the best models are listed in Table~\ref{tab:model}.

\subsection{HM~Sge}

The EPIC-pn spectrum of HM~Sge displays the obvious contribution from two components (see Fig.~\ref{fig:spec1} left panel). One dominating at energies below $E<0.5$~keV and a secondary contributing at $E>0.5$~keV. No significant emission is detected beyond 2.0~keV. 

The spectrum suggest at the presence of emission lines. 
The soft ($E<$0.5 keV) spectral region exhibits apparent peaks at 0.4, 0.3, and 0.2~keV with some emission in the 0.3--0.4 keV range. An additional peak is located at about $\gtrsim$0.7~keV. Given the low spectral resolution of the EPIC-pn spectra, the presence of emission lines might be questionable. However, 
the inspection of the 1st order RGS(1+2) spectrum of HM~Sge reveals the presence of several emission lines (see the top panel of Fig.~\ref{fig:RGS} in Appendix~\ref{sec:appendix}). The most prominent lines are those of S\,{\sc xi} 31.05,31.48~\AA\, ($\sim$0.4~keV), S\,{\sc xii} 32.4~\AA\, ($\sim$0.38~keV) and the unresolved N\,{\sc v} triplet at $\sim$29.0~\AA\,($\sim$0.43~keV). The presence of the C\,{\sc v} 35.0~\AA\,($\sim$0.35~keV), N\,{\sc v} 24.78~\AA\,(0.5~keV) and an unresolved contribution from the O\,{\sc v} 18.67~\AA\, and O\,{\sc vi} 18.97~\AA ($\sim$0.65~keV) can also be hinted in the spectrum.

Single-temperature plasma models were first attempted, but did not result in acceptable fits ($\chi_\mathrm{DoF}>4$). At best, these models only fit the spectrum for $E<0.5$~keV. The best model to the EPIC-pn spectrum of HM~Sge ($\chi_\mathrm{DoF}^2$=62.31/51=1.21) resulted in two-plasma components with temperatures of $kT_1=2.9\times10^{-2}$ keV (=$3.4\times10^{5}$~K) and $kT_2=0.31$~keV (=3.6$\times10^{6}$~K). The total intrinsic X-ray flux and luminosity for the 0.2--2.0~keV energy range are $F_\mathrm{X}$=(5.3$\pm$2.1)$\times10^{-10}$~erg~cm$^{-2}$~s$^{-1}$ and $L_\mathrm{X}$=$(6.3\pm2.5)\times10^{34}$~erg~s$^{-1}$, respectively. 
The detection of the hotter component is consistent with the analysis of the ROSAT PSPC observations presented in \citet{Murset1997}. However, the softer component was not detected in those ROSAT data even though the PSPC instrument had a higher sensitivity towards the soft energy range. We can only suggest that this emission has been enhanced in the recent years.

The extreme soft emission from symbiotic stars ($E<0.4$ keV) is usually attributed to the quasi-steady thermonuclear burning on the surface of the WD \citep{Orio2007}. However, in the case of HM~Sge it could be also explained by the presence of jets. In fact, \citet{Corradi1999} presented optical images to unveiled the presence of jet-like features at distances $\lesssim8''$ from HM~Sge. In addition, their optical spectra suggest deprojected jet velocities of $\sim$100~km~s$^{-1}$. Assuming an adiabatic shock, the  plasma temperature of $3.4\times10^{5}$~K suggests a shock velocity of $v_\mathrm{shock}\approx120 \mu^{-1/2}$~km~s$^{-1}$, where $\mu$ is the mean molecular weight of the particles. The latter is consistent with the estimated jet velocities. In addition, \citet{Goldman2022} presented evidence that HM~Sge has been dimming in the past recent years which they attribute to a change in the system orientation or to a sudden mass ejection.

It is interesting to note that the velocity of the jet in HM~Sge is at least an order of magnitude below the estimated escape velocity defined by a WD. Bipolar ejections can be produce by thermonuclear runaways at the surface of the WD \citep[see for example the case of the symbiotic recurrent nova RS~Oph;][]{Montez2022}, but the situation in this source might be different. \citet{AguayoOrtiz2019} presented their {\it chocked accretion} model where they explore the accretion of axi-symmetric, large-scale, small-amplitude in the density structures onto a gravitating object. They demonstrated that a natural consequence is the formation of bipolar outflows and, in particular, they found that for density contrasts between the equator and the poles of 0.1--1\,\%, the outflow has velocities of 0.1--0.3 times the local escape velocity.

HM~Sge might be similar to what has been reported for R~Aqr where the jet-like features are spatially resolved in soft X-ray emission by Chandra \citep{Kellogg2001} which are feeding an even more extended soft region detected by XMM-Newton \citep{Toala2022}. We note that in the case of HM~Sge the XMM-Newton observations do not resolve these jet-like features.

The contribution from the softer component ($E<0.5$ keV) questions previous classification of HM~Sge as a $\beta$-type symbiotic stars. It appears that a more accurate classification would be a $\alpha/\beta$-type, but we note that such classification has not been suggested in the literature \citep[see, e.g.,][and references therein]{Merc2019}. Finally, we note that such change in spectral characteristics raise strong questions on the initial classification of those sources that have not been monitored in recent years.

\subsection{PU~Vul}

The EPIC-pn spectrum of PU~Vul also seem to display the contribution from emission lines in the 0.2--2.0~keV (see Fig.~\ref{fig:spec1} right panel). One can attribute the dominant peaks in the spectrum to the contribution from N\,{\sc vi} 29.54~\AA\,(0.42~keV), N\,{\sc vi} 24.78~\AA\,(0.5~keV), and to a lesser extend to those of N\,{\sc vii} 19.36~\AA\,(0.64~keV). Some emission at $\sim$0.37~keV in the EPIC-pn spectrum might suggest at the presence of C\,{\sc vi} 33.7~\AA, but this wavelength range is noisy in the RGS spectrum of PU~Vul (see Fig.~\ref{fig:RGS} bottom panel).
A secondary peak between 0.8--1.0 keV can be attributed to the Fe\,{\sc xvii} 16.78~\AA\,(0.74~keV), Fe\,{\sc xvii} 15.02~\AA\,(0.83~keV), and Ne\,{\sc x} 12.14~\AA\,(1.0~keV) emission lines marginally detected in the RGS spectrum. No considerable emission is detected beyond 2.0~keV.

\begin{figure*}
\begin{center}
  \includegraphics[angle=0,width=0.4\linewidth]{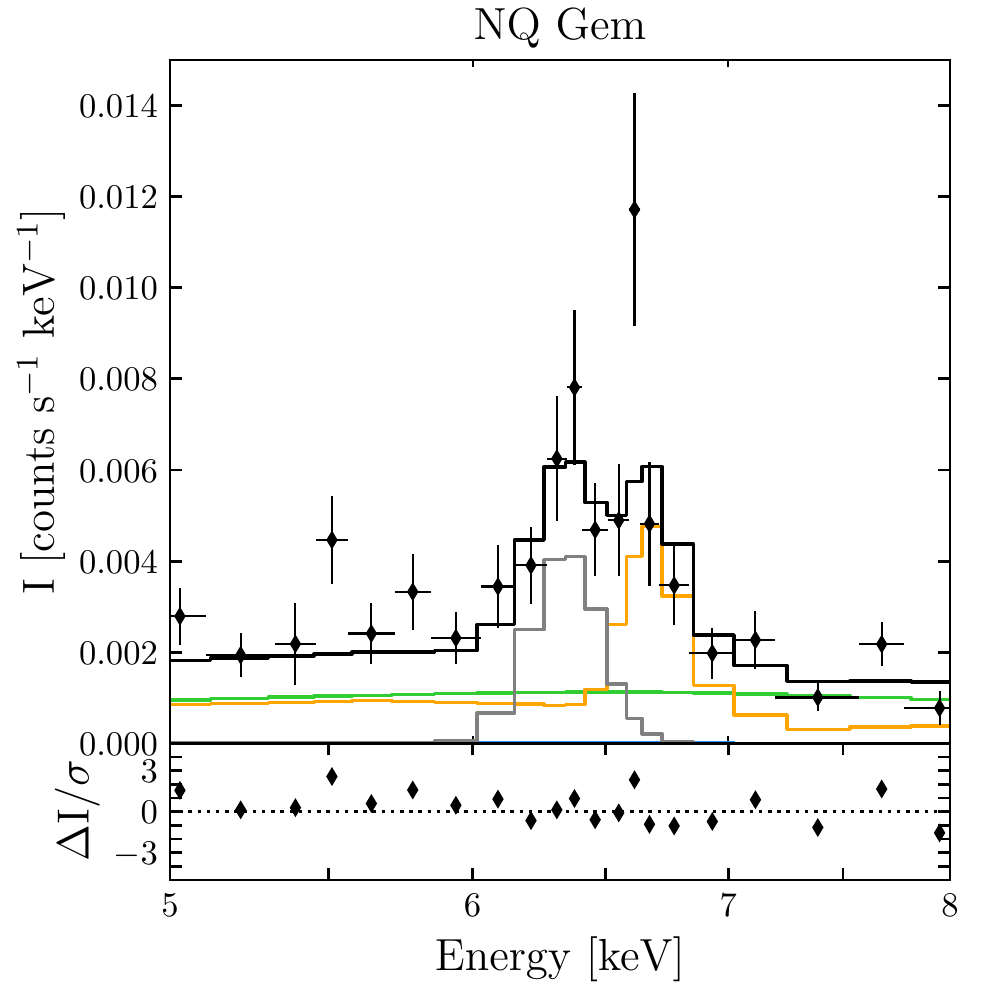}~
  \includegraphics[angle=0,width=0.4\linewidth]{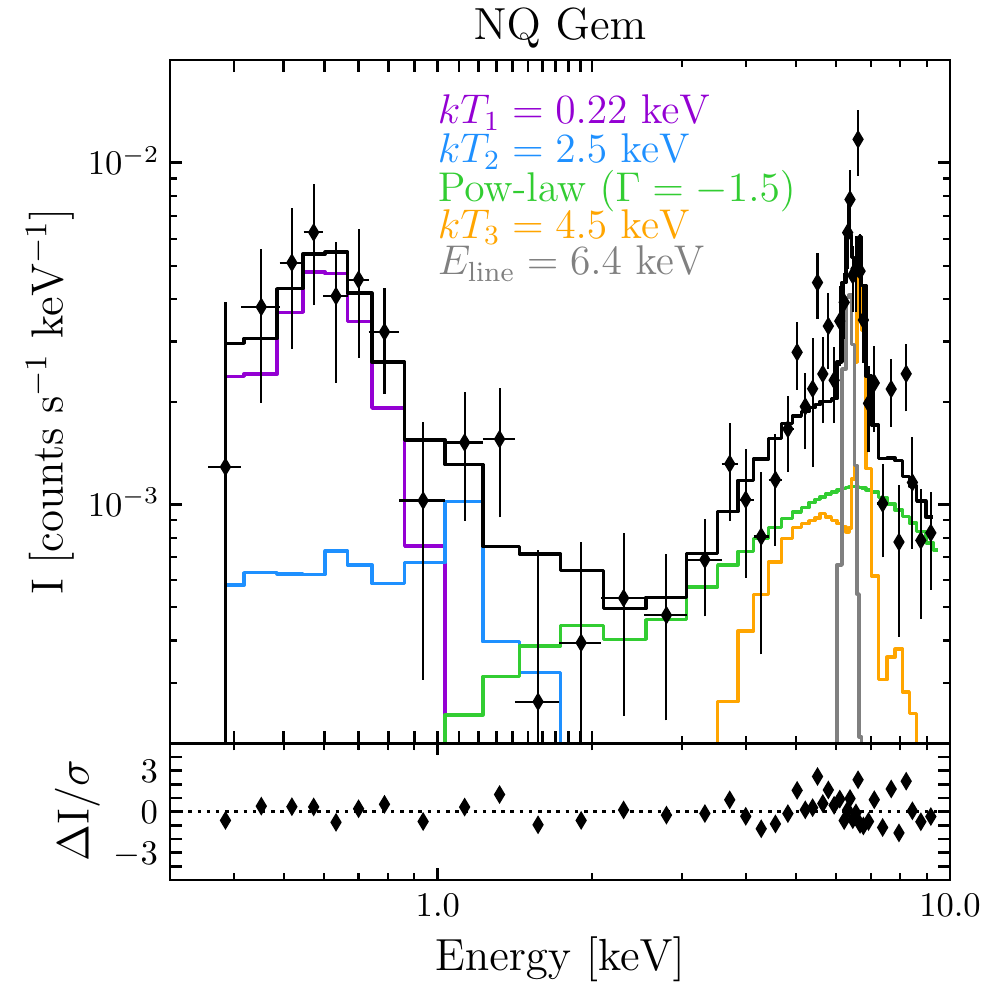}
\caption{Background-subtracted EPIC-pn spectrum of NQ~Gem (black diamonds). Left: Linear scale of the 5.0--8.0 keV energy range showing the Fe emission complex. Right: A log-linear scale of the EPIC-pn spectrum presented in the middle panel of Fig.~\ref{fig:spec1}. In both panels the best-fit model to the EPIC-pn spectrum is presented with a black histogram whilst different components are shown in colors (see Table~\ref{tab:model} for details). In both cases the bottom panels show the residuals.}
\label{fig:spec2}
\end{center}
\end{figure*}

Single-plasma temperature models did not result in acceptable fits to the EPIC-pn spectrum ($\chi_\mathrm{DoF}^2>2$). The best fit ($\chi_\mathrm{DoF}^2$=32.99/26=1.27) corresponds to a two-plasma model with typical temperatures of $kT_1$=0.15~keV (=$1.7\times10^{6}$~K) and $kT_2$=0.75~keV (=$8.7\times10^{6}$~K), with the softer component being the dominant one. The total intrinsic X-ray flux and luminosity are $F_\mathrm{X}$=(8.2$\pm$1.8)$\times10^{-14}$~erg~cm$^{-2}$~s$^{-1}$ and $L_\mathrm{X}$=$(1.6\pm0.4)\times10^{32}$~erg~s$^{-1}$.

\citet{Murset1997} obtained a relatively worse model to the ROSAT PSPC data. These authors restricted their study  to one-temperature plasma models given the quality of their spectrum. We note that the model presented for PU~Vul by \citet{Murset1997} does not  produce a good fit to the observed spectrum (see fig.~3 in that paper). Their dominant plasma temperature ($\sim6.2\times10^{6}$~K) represents an intermediate value of the temperatures obtained in our best model to the EPIC-pn spectrum. Our luminosity is also consistent (within error bars) with the one reported in \citet{Murset1997} once considering that those authors used a distance of 1.8~kpc to PU~Vul.

The EPIC data of PU~Vul presented here confirm this source as a $\beta$-type X-ray-emitting symbiotic star \citep{Murset1997}, where the strong wind from the WD component is slamming the slow wind from the cool component and producing X-ray-emitting gas. We note that a few authors \citep[see for example][]{Nussbaumer1996,Skopal2006} have reported that the WD component in PU~Vul has a stellar wind velocity of about $\gtrsim$1000~km~s$^{-1}$, which can easily produce shocked gas with temperatures of a few times 10$^{6}$~K.

\subsection{NQ~Gem}

The EPIC-pn spectrum of NQ~Gem discloses a more complicated spectral shape, as previously reported by \citet{Luna2013}. These authors classified NQ~Gem as a $\beta/\delta$-type symbiotic star given its clear presence of soft and hard emission in its spectrum. The EPIC-pn spectrum presented in the middle panel of Fig.~\ref{fig:spec1} reveals that the soft emission has a double peak, a broad dominant feature peaking t 0.55~keV, and a secondary peaking at $\sim$1.2~keV. The broad feature spans from 0.4 to 1.0~keV and seem to be the result of blending from different emission lines detected in the $\sim$13--30~\AA\, wavelength range of the RGS spectrum (see Fig.~\ref{fig:RGS} third panel). The secondary peak seems to include the contribution from emission lines in the $\sim$8--13~\AA\, (Fig.~\ref{fig:RGS} second panel). In the left panel of Fig.~\ref{fig:spec2} we show that the harder emission ($E>4.0$~keV) exhibits the presence of the Fe complex\footnote{This complex is composed by the fluorescent, He-like and H-like Fe lines at 6.4, 6.7 and 6.9~keV.} at $\sim$6.5~keV, the later marginally detected in the Swift data \citep[see fig.~1 in][]{Luna2013} and typically attributed to the presence of an accretion disk \citep[see, e.g.,][and references therein]{Eze2014}.

\citet{Luna2013} argued that the best fit to the Swift XRT spectrum of NQ~Gem includes a slightly-absorbed thermal component ($N_\mathrm{H,1}\lesssim$10$^{21}$~cm$^{-2}$, $kT_1$=0.23~keV) plus a heavily-absorbed component ($N_\mathrm{H,2}$=90$\times$10$^{21}$~cm$^{-2}$, $kT_2\gtrsim$16~keV). We started our spectral fitting by attempting similar models, but these were not able to appropriately reproduce the 1.0--4.0~keV energy range and did not result in acceptable fits ($\chi_\mathrm{DoF}^{2}>2$). In order to fit the secondary peak of the soft emission (that at $\sim$1.2~keV), it was necessary to include of a second {\it apec} component. However, that model did not result in an acceptable fit either. A subsequent model adding an extra component in order to fit the 2.0--4.0~keV energy range was required. Models including three slightly-absorbed thermal components did not result in acceptable fits. XSPEC was unable to restrict the plasma temperature of the third component, yielding extremely large values. Consequently, the best model ($\chi_\mathrm{DoF}^{2}$=36.20/32=1.13) includes two absorbed temperature models, a power law, and a heavily-absorbed thermal component with extra contribution from a Gaussian in order to fit the Fe emission. That is,
\begin{equation}
N_\mathrm{H,1}\times (apec_1+apec_2+pow) + N_\mathrm{H,2} \times (apec_3 + Gauss).
\end{equation}
\noindent Here the Gaussian component is needed to fit the 6.4~keV emission line. The parameters used in XSPEC to perform a fit using a Gaussian profile are the line energy $E_\mathrm{line}$, the line width ($\sigma_\mathrm{line}$) in keV, and normalization parameter ($A_\mathrm{line}$).

The soft energy range range of the EPIC-pn spectrum ($E<4.0$ keV) is reproduced by the contribution from plasma components with temperatures of $kT_1$=0.22~keV and $kT_2$=0.31 keV in addition to a power law with a photon index of $\Gamma=-1.25$ that also contributes to the harder spectral region (see Fig.~\ref{fig:spec2} right panel). The heavily-absorbed component has an hydrogen column density of $N_\mathrm{H,2}=3.9\times10^{23}$~cm$^{-2}$ with a plasma temperature of $kT_3$=4.5~keV. This component also produces the contribution from the He-like component of the Fe emission line at 6.7~keV (orange line in Fig.~\ref{fig:spec2}). The total intrinsic X-ray flux and luminosity of this model are $F_\mathrm{X}$=(2.7$\pm$0.5)$\times10^{-13}$~erg~cm$^{-2}$~s$^{-1}$ and $L_\mathrm{X}$=$(3.4\pm0.6)\times10^{31}$~erg~s$^{-1}$, respectively. These values are slightly smaller than those presented by \citet{Luna2013}.

X-ray spectral modeling of $\beta/\delta$-type symbiotic stars is usually attempted with simpler models as that presented here. But the higher quality EPIC-pn spectrum of NQ~Gem suggests that some extra components are to be taken into account to appropriately fit the 1.0--4.0~keV energy range. We note that the power law component (usually associated with non-thermal emission) might not be the best option to assess the physics behind the production of X-rays in NQ~Gem. The power law component improves the goodness of the fit, but it might suggest the need of a continuous distribution of temperatures.

This situation might be the same for other other $\beta/\delta$-type symbiotic stars, such is the case of CH~Cyg. For example, \citet{Mukai2007} presented Suzaku observations of CH~Cyg and used a two-plasma components fits to model the XIS spectra. Their figure 1 shows that the 2.0--4.0~keV energy range is not very well fitted by the proposed model. The analysis of archival XMM-Newton observations of CH~Cyg suggest that another approach to fit this energy range is to include the presence of a reflection component \citep{Ishida2009} very similar to that used for active galactic nuclei (Toal\'{a} et al. in prep.). A subsequent study of high-quality observations of $\beta/\delta$-type symbiotic stars will help understanding accretion processes and the effects of reflection behind the production of X-rays.

\section{SUMMARY}   
\label{sec:summary}

We presented the analysis of archival XMM-Newton observations of three symbiotic stars, namely, HM~Sge, NQ~Gem, and PU~Vul. In the three cases, their EPIC-pn spectra have higher-quality than previous spectra presented in the literature. The EPIC-pn spectra hint at the presence of emission lines that is further corroborated by the RGS spectra of the three sources. The analysis of these sources improves previous determination of their X-ray properties, but the determination of fluxes and luminosities agree with previous works. Our findings can be summarized as follows:
\begin{itemize}

\item \underline{HM~Sge}. We detected an extra soft component in the EPIC-pn spectrum of HM~Sge not detected by ROSAT, similarly to what is found for $\alpha$-type symbiotic stars. This corresponds to a plasma with temperature of 2.9$\times10^{-2}$~keV (=3.4$\times10^{5}$~K). Although $\alpha$-type symbiotic stars are associated with thermonuclear burning on the surface of the WD component, we suggest that in the case of HM~Sge it might be attributed to the action of jets with velocities of $\gtrsim$100~km~s$^{-1}$. We suggest a spectral classification of $\alpha/\beta$ which is an new alternative to the one proposed before. 

\item \underline{PU~Vul}. The model that best reproduces the EPIC-pn spectrum of PU~Vul includes the contribution from two thermal plasma components, 0.22~keV (=1.7$\times10^{6}$~K) and 0.75~keV (=8.7$\times10^{6}$~K). We confirm that this symbiotic star can be classified as a $\beta$-type.

\item \underline{NQ~Gem}. The EPIC-pn spectrum of this $\beta/\delta$-type symbiotic star unambiguously exhibits the presence of the Fe emission lines above 6.0~keV. Two-temperature plasma models, which are typically used to fit the spectrum of this class of symbiotic star, fail to reproduce the EPIC-pn spectrum of NQ~Gem. The model requires at least two extra components in order to fit the 1.0--4.0~keV energy range: an extra optically-thin plasma and a power law component. We suggest that the power law component, usually associated with non-thermal emission does not represent the true nature of the X-ray emission in this source. Alternatively, we suggest that more sophisticated models including reflection components should be used to model $\beta/\delta$-type symbiotic stars.

\end{itemize}

\begin{acknowledgments}
The authors thank comments and suggestions from an anonymous referee that helped improving the presentation and interpretation of the results. J.A.T. thanks Fundaci\'{o}n Marcos Moshinsky (Mexico) and the UNAM PAPIIT project IA101622. M.K.B. thanks Consejo Nacional de Ciencias y Tecnolog\'{i}a (CONACyT, Mexico) for research student grant. L.S. and M.K.B. also acknowledge support from UNAM PAPIIT project IN110122. This work is based on observations obtained with XMM-Newton, an European Science Agency (ESA) science mission with instruments and contributions directly funded by ESA Member States and NASA. This work has made extensive use of NASA's Astrophysics Data System.
\end{acknowledgments}

%

\vspace{5mm}
\facilities{XMM-Newton\,(EPIC)}


\software{SAS \citep{Gabriel_2004}, XSPEC \citep{Arnaud1996}}



\appendix
\setcounter{figure}{0}
\renewcommand{\figurename}{Figure}
\renewcommand{\thefigure}{A\arabic{figure}}

\section{RGS spectra}
\label{sec:appendix}

The RGS spectra of the three symbiotic stars analyzed here were processed using the SAS tasks {\it rgsproc}. This task produces source and background spectra as well as the necessary calibration matrices. First order spectra were extracted from the RGS1 and RGS2 instruments and were then combined using the SAS task {\it rgscombine}.

\begin{figure*}
\begin{center}
  \includegraphics[angle=0,width=0.8\linewidth]{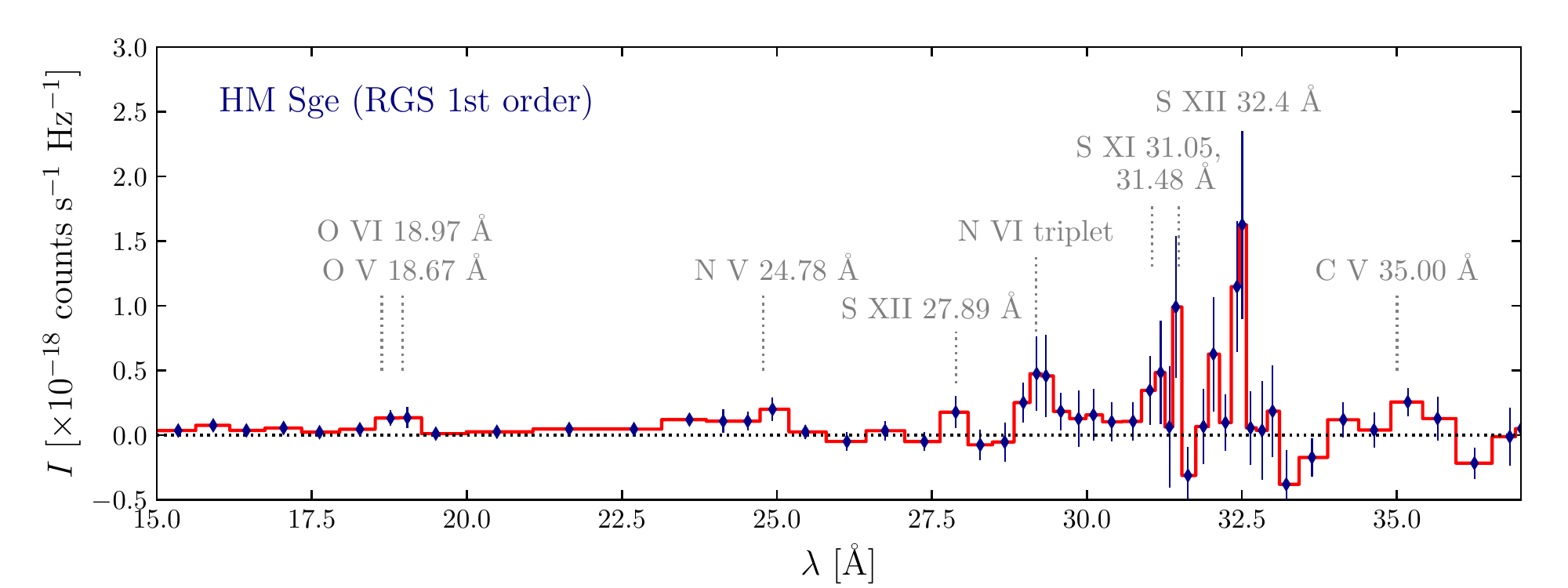}\\
  \includegraphics[angle=0,width=0.8\linewidth]{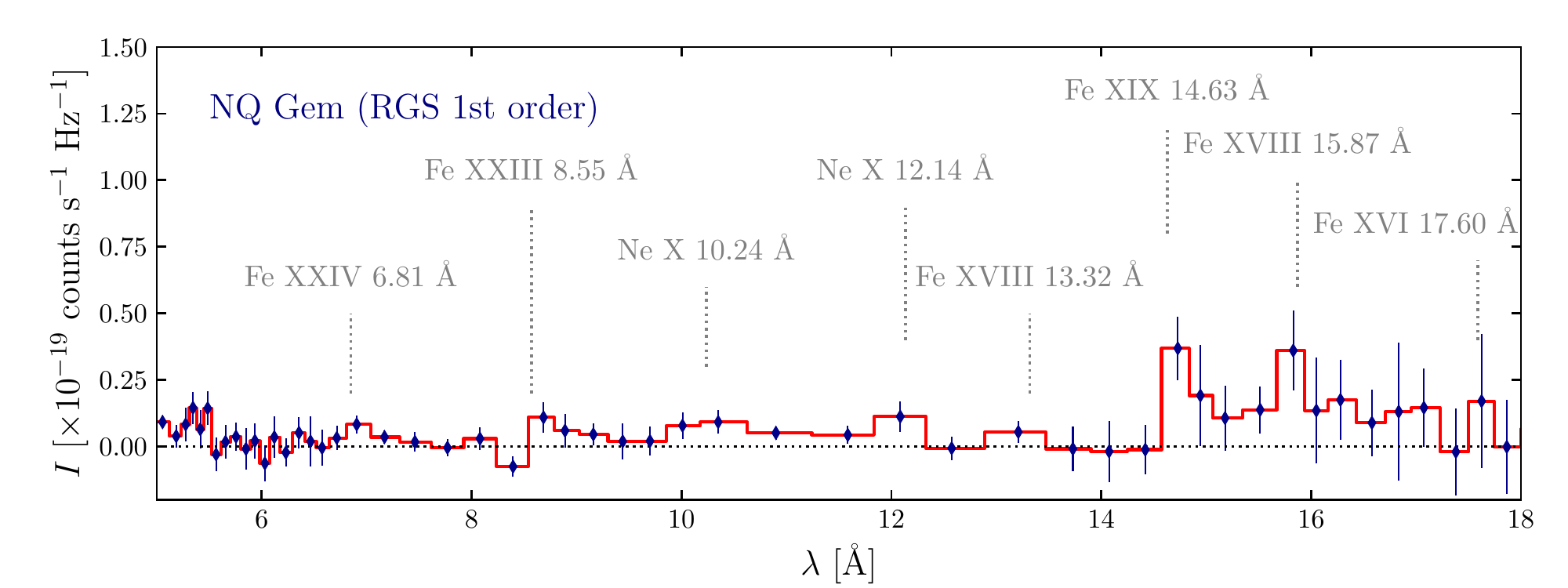}\\
  \includegraphics[angle=0,width=0.8\linewidth]{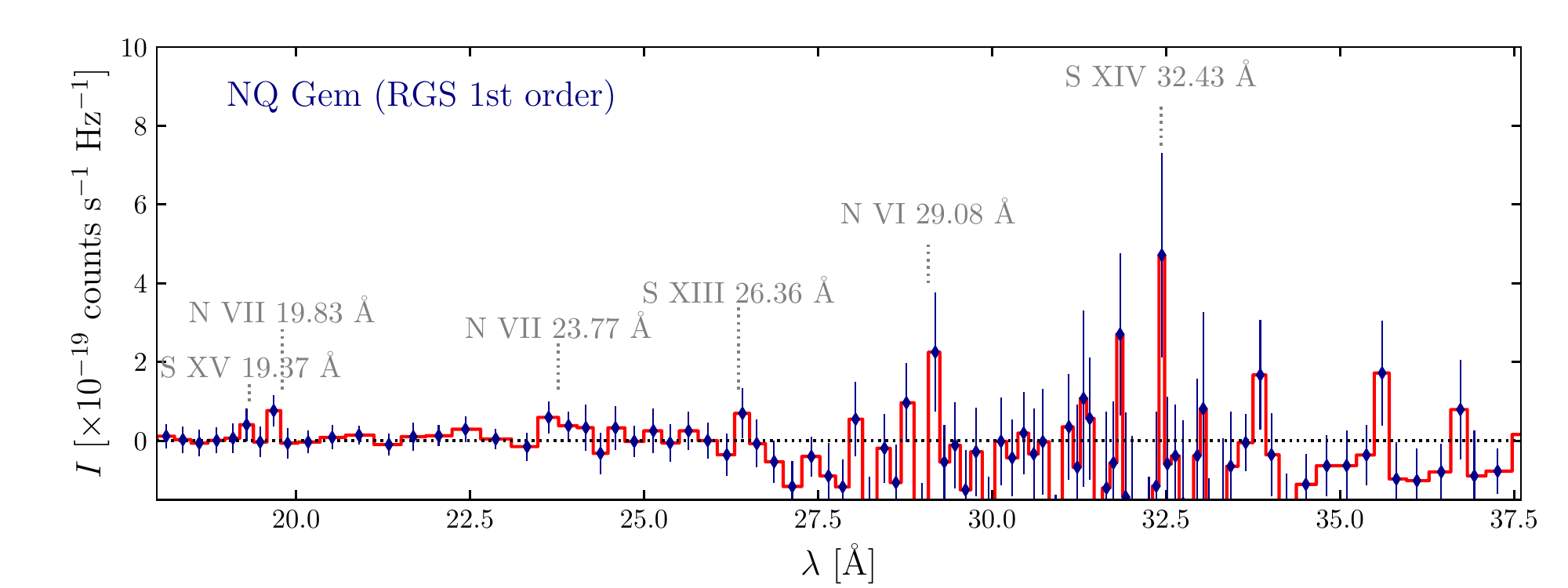}\\
  \includegraphics[angle=0,width=0.8\linewidth]{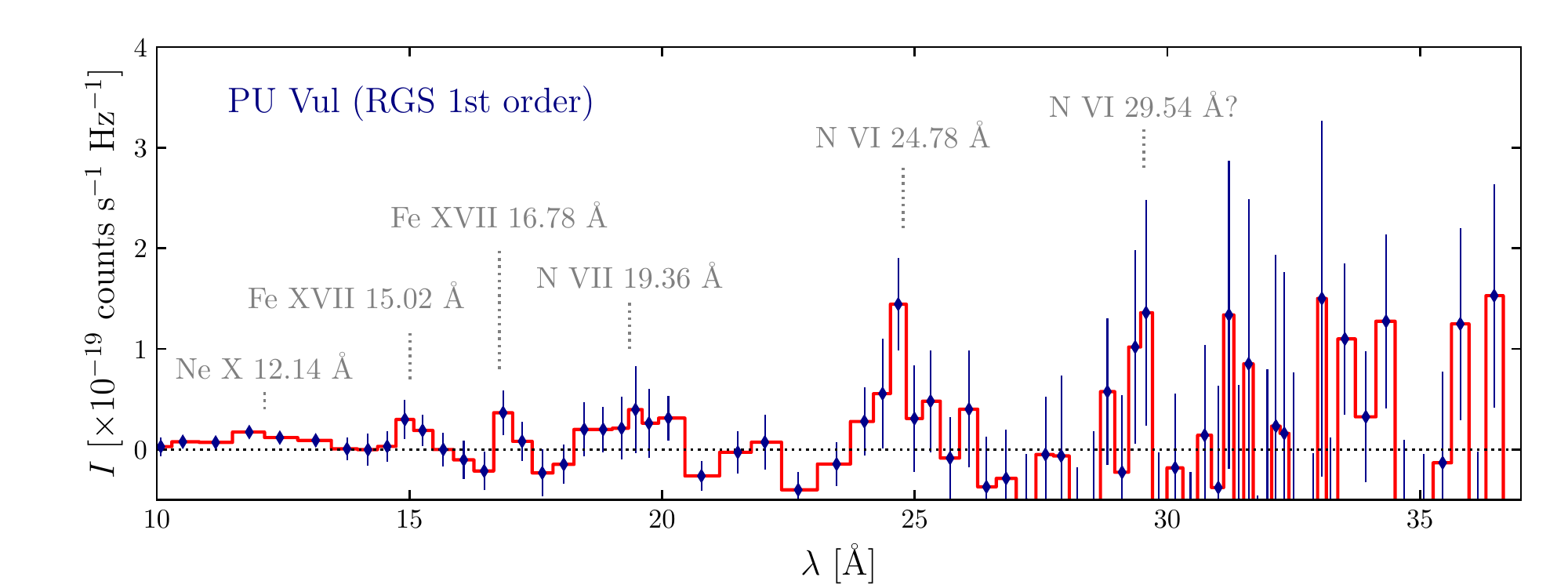}  
\caption{1st order RGS(1+2) spectra of HM\,Sge (first panel), NQ~Gem (second and third panels), and PU~Vul (bottom panel). The most prominent lines are labeled.}
\label{fig:RGS}
\end{center}
\end{figure*}

In Fig.~\ref{fig:RGS} we present the background-subtracted 1st order RGS(1+2) spectra of HM~Sge, NQ~Gem, and PU~Vul. The RGS spectra cover the 5--38~\AA\, wavelength range which corresponds to the $\sim$0.32--2.48~keV energy range. The most prominent emission lines are labeled on each spectrum. Although shallow, these spectra confirm that the peaks in the EPIC-pn spectrum can be related to the presence of emission lines. A detailed analysis of RGS spectra of a sample of symbiotic stars will be presented in a separate work.

\end{document}